\documentstyle[twocolumn,seceq]{jpsj}
\renewcommand{\theequation}{\arabic{equation}}
\catcode`\@=11
\def\simle{\mathrel{\mathpalette\@versim<}}   
\def\simge{\mathrel{\mathpalette\@versim>}}   
\def\@versim#1#2{\lower2.5pt\vbox{\baselineskip0pt \lineskip-.5pt
   \ialign{$\m@th#1\hfil##\hfil$\crcr#2\crcr\sim\crcr}}}
\catcode`\@=12
\def\runtitle{}
\def\runauthor{}

\title
{
Effects of Orbital Degeneracy and Electron Correlation  \\ 
on Charge Dynamics in Perovskite Manganese Oxides
}

\author
{ 
Hiroki {\sc Nakano}, Yukitoshi {\sc Motome}$^{1}$ and 
Masatoshi {\sc Imada}
}

\inst
{
Institute for Solid State Physics, University of Tokyo, 
Roppongi 7-22-1, Minato-ku, Tokyo 106-8666\\
$^1$Department of Physics, Tokyo Institute of Technology, Oh-okayama 
2-12-1, Meguro-ku, Tokyo 152-8551
}

\recdate
{
}

\abst
{
Taking the orbital degeneracy of $e_g$ conduction bands and 
the Coulomb interaction into account in a double-exchange model, 
we investigate charge dynamics of perovskite Mn oxides 
by the Lancz$\ddot{\rm o}$s diagonalization method. 
In the metallic phase near the Mott insulator, 
it is found that the optical conductivity 
for a spin-polarized two-dimensional system 
exhibits a weight transfer to a broad and incoherent structure 
within the lower-Hubbard band together 
with a suppressed Drude weight. 
It reproduces qualitative feature of the experimental results.  
As an orbital effect, we find that an anomalous charge correlation 
at quarter filling suppresses the coherent charge dynamics 
and signals precursor to the charge ordering.  
}

\kword
{
perovskite manganites, double-exchange model, Coulumb interaction, 
orbital degeneracy, metal-insulator transition, ferromagnetic metal, 
optical conductivity, Drude weight, incoherent charge dynamics, 
charge ordering, 
exact diagonalization, strongly correlated electron system
}

\begin{document}
\sloppy
\hyphenation{Hamil-ton-ian}
\maketitle

Although properties of perovskite manganese oxides 
$R_{1-x}A_{x}$MnO$_3$ ($R$=rare earth, $A$=Ca, Sr, Ba or Pb) 
showing colossal magnetoresistance\cite{Ramirez_rev} 
are to some extent reproduced by a double-exchange (DE) model 
composed of non-interacting (NI) electrons 
in a non-degenerate $e_{g}$ conduction band 
ferromagnetically coupled with $t_{2g}$ localized spins 
of $S$=3/2\cite{Zener,Anderson_Hasegawa,de_Gennes,Kubo_Ohata,Frkw_DMF}, 
there remain open problems which cannot be understood within
the framework of this model.  
One of them is that 
La$_{1-x}$Sr$_{x}$MnO$_3$ 
in the ferromagnetic and metallic phase near the insulator ($x$=0.175) shows 
a large and broad structure of incoherence 
in the optical response together with a relatively small 
Drude weight at low temperatures\cite{Okimoto_let,Okimoto_reg} 
though the spin degrees of freedom cannot contribute 
to the incoherence under the perfect polarization of spins.  
Some other degrees of freedom not contained in the DE model 
must be involved.  
Possible origins of the incoherence are 
the dynamical Jahn-Teller (JT) distortion\cite{Millis}, 
$e_{g}$ orbital 
degeneracy\cite{Shiba_Shiina_Takahashi,Takahashi_Shiba,Brito_Shiba} and 
electronic correlations\cite{Ishihara_Yamanaka_Nagaosa,Horsch_finite_temp}. 
In ref. \ref{Millis} employing the dynamical mean-field approximation, 
the electronic correlations are not taken into account 
though they are not negligible and 
play an essential role as shown later. 
In this work, we focus on the latter two 
and make it clear how large incoherence is induced 
in a minimal model of these two origins. 

On this subject, related studies have already 
been done under some approximations. 
Shiba, Shiina and Takahashi claimed that 
the large incoherence in the optical absorption 
originates in the degeneracy of the $e_{g}$ 
orbitals\cite{Shiba_Shiina_Takahashi},   
although the Mott insulator in the undoped system is not 
reproduced because the strong correlation effect is not seriously 
considered\cite{Shiba_Shiina_Takahashi,Takahashi_Shiba,Brito_Shiba}.  
Ishihara, Yamanaka and Nagaosa 
derived an effective $t$-$J$ type Hamiltonian 
considering the degeneracy of the $e_{g}$ orbitals 
and the strong interaction\cite{Ishihara_Yamanaka_Nagaosa}. 
They obtained incoherent part of 
the optical conductivity in a mean-field treatment. 
In such a mean-field treatment, unfortunately, 
the Drude weight is absent 
with a diverging specific heat coefficient $\gamma$ 
in contrast to the experimental results\cite{Woodfield}.  
Horsch {\it et al.} applied a finite-temperature Lancz$\ddot{\rm o}$s 
method to the `orbital' $t$-$J$ model\cite{Horsch_finite_temp}.  
In their method, however, the accessible temperature range is much higher 
than the experimental conditions\cite{Horsch_temperature}. Consequently 
their Drude weight is largely suppressed simply by thermal fluctuations, 
which makes hard to compare with the experimental results.  
In order to discuss this metal-insulator transition, 
fluctuation effects are important; a reliable way to 
calculate beyond biased approximations is required.

 In this work, 
we examine the Hamiltonian given by 
\begin{equation}
{\textstyle 
{\cal H}\hspace{-1mm}=\hspace{-1mm}\sum_{ij}\sum_{\nu\nu^{\tiny \prime}}
t_{ij}^{\nu\nu^{\tiny \prime}} c^{\dagger}_{i\nu} 
c_{j\nu^{\tiny \prime}} + U \sum_{i}(n_{i1}-\frac{1}{2})(n_{i2}-\frac{1}{2}), 
}
\label{Mn-Hamiltonian}
\end{equation}
as a minimal model mentioned above. 
This model is derived from the generalized DE model with 
$e_{g}$-orbital degeneracy under the assumptions 
of strong Hund's-rule coupling and perfect spin 
polarization\cite{Motome_Imada}. 
Here $t_{ij}^{\nu\nu^{\prime}}$ denotes the hopping integral 
and $U$ is the effective interorbital Coulomb interaction 
obtained after subtracting the Hund's-rule coupling energy 
between $e_{g}$ electrons. 
The orbitals $d_{x^2-y^2}$ and $d_{3 z^2-r^2}$ correspond to 
$\nu$\mbox{\,}$=$\mbox{\,}$1$ and  $\nu$\mbox{\,}$=$\mbox{\,}$2$, 
respectively. 
Note that this model is essentially different 
from the usual Hubbard model. Actually, the difference appears 
even in the NI case as shown later. 
We consider here only the nearest-neighbor hopping given by 
$t_{ij}^{11}=-3/4 t_0$, $
t_{ij}^{22}=-1/4 t_0$, $
t_{ij}^{12}$\mbox{\,}$=$\mbox{\,}$t_{ij}^{21}$\mbox{\,}$=$\mbox{\,}$
-(+)\sqrt{3}/4 t_0$ along
$x$($y$)-direction in two dimensions. 
Here, we investigate the two-dimensional (2D) model given 
by eq.$\mbox{\,}$(\ref{Mn-Hamiltonian}) based on the following reasons. 
One is that the three-dimensional (3D) system of 
$R_{1-x}A_{x}$MnO$_{3}$ keeps a perfectly spin-polarized 
plane not only in the ferromagnetic doped phase 
but also in the $A$-type antiferromagnetic insulating phase 
at $x$=0 so that the spin polarization is retained 
in a 2D plane over all regions of interest. 
The second is that 
a similar incoherent feature in the optical conductivity was also 
reported in the 2D system, for instance, 
in La$_{1.2}$Sr$_{1.8}$Mn$_{2}$O$_{7}$\cite{Ishikawa}.  
It is therefore likely that the 2D system can capture 
basic physics of the incoherent charge dynamics. 
The model (\ref{Mn-Hamiltonian}) contains an implicit 
variable parameter of a doping concentration 
$\delta$\mbox{\,}$\equiv$\mbox{\,}$1-\frac{N_{\rm e}}{N_{\rm s}}$ or 
an electron density $n\equiv \frac{N_{\rm e}}{N_{\rm s}}$, 
where $N_{\rm e}$ and $N_{\rm s}$ denote the numbers of electrons and 
sites, respectively. 
Recently, this model (\ref{Mn-Hamiltonian}) has been studied 
by means of quantum Monte Carlo (QMC) 
method\cite{Motome_Imada,NEC_Motome_Nakano_Imada,Motome_reg}. 
There, the relation was discussed between 
the incoherence of the charge dynamics and 
divergences of both a staggered orbital correlation length 
and the compressibility as $\delta$\mbox{\,}$\rightarrow$\mbox{\,}0. 
To get more insight, it is desired to calculate quantities directly 
connected to the charge dynamics. 
In this paper, then, we discuss 
effects of the orbital degeneracy and the electronic correlation 
based on direct calculations of dynamical quantities 
about charge transport.  

 We here employ, as another unbiased method, 
exact diagonalization of finite-size (FS) clusters, 
using the Lancz$\ddot{\rm o}$s algorithm and 
the continued-fraction-expansion method\cite{frac_expansion} 
to obtain dynamical quantities at zero temperature. 
The sizes of the cluster calculated in this work are 
$\sqrt{10} \times\!\sqrt{10}$ and $4\times 4$. 
In the $\sqrt{10} \times\!\sqrt{10}$-site system,  
to reduce the FS effect\cite{Nakano_Imada_usual-Hub}, 
we introduce the Aharonov-Bohm flux $\Phi$ 
in the hopping term as the twisted boundary condition 
and find the optimized boundary so as to minimize 
the ground-state energy. 
In the $4\times 4$ system, we choose the boundary condition 
among the periodic ($\Phi = (0,0)$), the anti-periodic 
($\Phi = (\pi,\pi)$) and the mixed ($\Phi = (0,\pi)$ or $(\pi,0)$) ones.  

  We here calculate the optical conductivity defined as
$\sigma(\omega)$=[$\sigma_{x} (\omega)$+$\sigma_{y} (\omega)$]/2 
where  
$
\sigma_{\alpha} (\omega) = 2\pi e^2 D_{\alpha} \delta(\omega) 
+\frac{\pi e^2}{N}
{
\sum_{m(\ne 0)}}
\frac{|\langle m|j_{\alpha}|0\rangle |^{2}}{E_m-E_0} 
\delta(\omega -E_m+E_0)
$. 
Here, $D_{\alpha}$ is the Drude weight; $j_{\alpha}$ is a 
current operator along $\alpha$-direction 
($\alpha$={\it x},\mbox{\,}{\it y})  given by 
$
j_{\alpha} = - {\rm i}\sum 
t_{i,i+\delta_{\alpha}}^{\nu\nu^{\prime}}
(
 c^{\dagger}_{i\nu}                     c_{i+\delta_{\alpha} ,\nu^{\prime}}
- c^{\dagger}_{i+\delta_{\alpha} ,\nu^{\prime}} c_{i\nu}
) 
$; 
and $|m\rangle$ 
denotes an eigenstate of the sysytem with the energy eigenvalue 
of $E_m$. Note that $m=0$ represents the ground state. 
The averaging operation is performed to reduce the FS effect and 
to cancel the anisotropy due to the anisotropic boundary condition.  
The sum rule 
$
\int_{0}^{\infty} \sigma (\omega) d \omega = \pi e^2 K
$ 
is satisfied, where $-4K$ is the kinetic energy per site. 
We will call $K$ the total weight hereafter.  
When $U$ is large enough, we also calculate 
the effective carrier density defined as 
$
N_{\rm eff} = 
\frac{1}{\pi e^2 } 
\int_{0}^{\omega_{\rm c}}
\sigma (\omega) d\omega 
$
where $\omega_{\rm c}$ is a frequency just below the bottom edge 
of the upper-Hubbard (UH) band. 
In this work, we make a further procedure 
to reduce FS effects. 
The procedure is a multiplication of a correction factor 
determined from the FS effects in the NI cases. 
The factor is defined by $r=K_{\infty}^{(U=0)}/K_{N_{\rm s}}^{(U=0)}$, where 
$K_{\infty}^{(U=0)}$ and $K_{N_{\rm s}}^{(U=0)}$ are the total weights 
in the thermodynamic limit and in the $N_{\rm s}$-site case 
when $U$=0, respectively. 
Hereafter, all the quantities obtained after this procedure 
are labeled with suffix c, for example, $D^{\rm c}$. 
This procedure successfully reduces 
the FS effects in the usual Hubbard model 
in one and two dimensions\cite{Nakano_Imada_usual-Hub}. 
Here, we take a large value of $U/t_0$=16, 
at which a large Hubbard gap 
opens at half filling and the FS effect 
is expected to be less serious.


We first show our results of the doping dependence of the total weights, 
the effective carrier densities and the Drude weights 
in Fig. \ref{fig1}.   
Even in the NI case, 
as was reported in the 3D model 
in ref. \ref{Shiba_Shiina_Takahashi}, 
the present model (\ref{Mn-Hamiltonian}) in 2D  
exhibits a different doping dependence 
between the Drude weight and the total weight 
due to the hopping between the $d_{x^2-y^2}$ and $d_{3 z^2-r^2}$ bands. 
(See the gray curves in Fig. \ref{fig1}.)
However, the Drude weight remains finite 
even at half filling, which indicates that the system is metallic 
in this NI case. 
When $U/t_0 = 16$, on the other hand, 
the Drude weight at half filling 
has a very small value $\sim 0.0005$
in the case of $\sqrt{10} \times\!\sqrt{10}$-site system.  
This small value indicates that the system is Mott insulating and 
that even a cluster of $\sqrt{10} \times\!\sqrt{10}$ sites 
can reproduce well the bulk quantities. 
In the dilute-electron-density region ($\delta \sim 1$), 
the total weights, the effective carrier density and the Drude weights 
for $U/t_0 =16$ behave close to the corresponding quantities 
for the NI case, which indicates that effects of Coulomb 
interaction are small in this region. 
At $\delta = 0.5$, the Drude weight shows a significant dip, 
which reminds us of a charge ordering 
with a $CE$-type magnetic structure observed in such materials as  
Nd$_{0.5}$Sr$_{0.5}$MnO$_3$\cite{Kuwahara} 
and  Pr$_{0.5}$Ca$_{0.5}$MnO$_3$\cite{Jirak,Tomioka}.  
Although the value of the Drude weight is still finite, 
this dip is neither seen in the NI case of the present model nor 
in the usual Hubbard model with finite $U$\cite{Nakano_Imada_usual-Hub}. 
We will discuss this issue later. 
With decreasing $\delta$ further, 
the Drude weight and the effective carrier density vanish 
for $U/t_0$=16 while the two quantities increase for $U$=0.  
It should be emphasized here that the overall behaviors 
of the Drude weight and the effective carrier density 
except for the dip at $\delta$=0.5 in the Drude weight 
are qualitatively similar to the ones of the usual 
Hubbard model in 2D\cite{Nakano_Imada_usual-Hub}.  
We, however, also note that the critical region 
of suppressed Drude weight near $\delta$=0 seems to be narrower 
than that in the 2D Hubbard model 
due to the imbalanced populations of electrons 
in $d_{x^2-y^2}$ and $d_{3z^2-r^2}$. 

   In Fig. \ref{fig2}, we present results 
for the incoherent part of $\sigma (\omega )$. 
One can see in Fig. \ref{fig2} (a) that, at half filling, 
a large gap exists and that only the weight transfer across the gap 
to the UH band appears. 
As shown in Figs. \ref{fig2} (b) and (c), 
the more holes from half filling 
are doped, the more weights are transferred 
from the region above the gap 
to the region within the lower-Hubbard (LH) band. 
Especially, note that weights for $U/t_0$=16 spread 
in a wider region than those in the NI case 
shown in the inset of Fig. \ref{fig2}(b).  
The QMC study shows a critical enhancement 
of the orbital correlation length induced by $U$ 
which would cause the incoherence 
of the charge dynamics\cite{Motome_Imada}. 
The orbital correlation in the present calculation is 
in agreement with that in the QMC. 
The present calculations indeed 
show a close relation between the orbital fluctuation 
and the mid-gap incoherence. 
The qualitative feature of the incoherent charge response 
experimentally observed in $\sigma (\omega)$ is 
reproduced in the present calculation, 
which supports the importance of combined effects from orbital
degeneracy and strong correlations. 
On a quantitative level, however, 
$D/K$\mbox{\,}$\sim$0.51 at $\delta$\mbox{\,}$\sim$0.125 
in the present result is still larger 
than the experimental indications, $\sim$0.2. 


  To understand the dip at $\delta$\mbox{\,}$=$\mbox{\,}$0.5$ 
in Fig.\mbox{\,}\ref{fig1}, 
let us discuss the orbital and charge structures 
in the ground state at $\delta$\mbox{\,}$=$\mbox{\,}$0.5$.  
\cite{comment_quarter_filling_boundary}
The basic point is that an orbital-po\-larized state with 
a spatially anisotropic overlap seems to 
enhance the charge ordering as we see below. 
To gain the kinetic energy 
in such a state at finite $U$ and $\delta$\mbox{\,}$=$\mbox{\,}$0.5$,  
an occupied site favors a vacant site 
as the nearest neighbors. 
In the model (\ref{Mn-Hamiltonian}), 
the orbital polarization (OP) 
defined by 
$(\langle n_{x^2-y^2}\rangle 
-\langle n_{3z^2-r^2}\rangle)/(\langle n_{x^2-y^2}\rangle
+\langle n_{3z^2-r^2}\rangle)$ 
grows 
because $d_{x^2 - y^2}$ orbital has larger
hopping matrix than $d_{3 z^2 - r^2}$ orbital. 
The strong interaction also enhances the polarization\cite{Motome_Imada}.  
Actually the OP is 
$\sim$\mbox{\,}0.756 at $\delta$\mbox{\,}$=$\mbox{\,}$0.5$ 
for $U/t_0$=16\cite{comment_free_case} 
in the 4$\times$4-site system. 
Note here that, on the other hand, the usual Hubbard model 
does not show such a spin polarization 
in the ground state at $\delta$\mbox{\,}$=$\mbox{\,}$0.5$.  
When one uses $|\theta\rangle$\mbox{\,}$=$ $ \cos \theta 
| d_{x^2-y^2} \rangle + 
\sin \theta | d_{3z^2-r^2} \rangle$ as a notation 
for a single-site state, the above polarization leads 
to $|\theta|/\pi \simeq 0.114$, 
where each occupied orbital is strongly anisotropic and 
has a large overlap only in the $x$- or $y$-direction.  
We have next calculated the charge density correlation defined as 
$
\Delta (\mbox{\boldmath $k$})$\mbox{\,}$=$\mbox{}$ 
N_{\rm s}^{-1}$\mbox{}$
\sum_{i,j} 
   [ \langle n_{i}               n_{j}  \rangle 
  $\mbox{\,}$-$\mbox{\,}$ \langle n_i \rangle \langle n_{j} \rangle ] 
      \exp [ {{\rm i} \mbox{\boldmath $k$} \cdot 
    (\mbox{\boldmath $x$}_i-\mbox{\boldmath $x$}_{j})} ]
$
in the 4\mbox{\,}$\times$\mbox{\,}4\mbox{\,} sites, where 
$n_{i}$\mbox{\,}$\equiv$\mbox{\,}$n_{i1}$\mbox{\,}$+$\mbox{\,}$n_{i2}$.  
The results at $\delta = 0.5$ show a peak 
at $\mbox{\boldmath $k$}$\mbox{\,}$=$\mbox{\,}$(\pi,\pi)$ 
which makes staggered charge cor\-relation (CC). 
The doping dependence of the above CC 
at ($\pi ,\pi$) is shown in Fig.\mbox{\,}\ref{fig3}. 
An anomalous peak appears at $\delta$\mbox{\,}$=$\mbox{\,}$0.5$ 
only for the model (\ref{Mn-Hamiltonian}) while the usual Hubbard model 
exhibits a monotonic $\delta$ dependence. 
A similar behavior at $\delta$\mbox{\,}$=$\mbox{\,}$0.5$ 
of the CC 
and the charge dynamics was reported in ref. \ref{Ishihara_CO}, 
where a spinless and orbitless system 
with the usual Hubbard-type hoppings and 
the nearest-neighbor interaction $V$ was studied.  
Although the size treated here is too small to judge whether 
the present system is metallic or insulating, 
an anomalous behavior of $\Delta (\pi,\pi)$ 
at $\delta$\mbox{\,}$=$\mbox{\,}$0.5$ occurs 
even in the present case without $V$.  
The enhanced CC at $\delta$\mbox{\,}$=$\mbox{\,}$0.5$ 
which makes the suppressed coherence in the charge dynamics can be induced 
only by the anisotropic hopping and the on-site interaction 
while the static charge order will be more stabilized 
with the help of the intersite Coulomb replusion $V$. 
The JT distortion, studied theoretically by means 
of LDA+$U$\cite{Anisimov} and Hartree-Fock\cite{Mizokawa} methods, 
will also enhance the above tendency  
of the CC and the incoherence of the charge dynamics. 
Although three dimensionality would reduce the OP, which influences 
the stability of this staggered CC, 
our results in 2D system are important 
because the charge-ordering 
phenomena essentially appear with strong 2D anisotropy 
in experiments\cite{Kuwahara,Jirak,Tomioka,murakami}. 
To obtain information of orbital patterns at $\delta$=0.5, in addition, 
we calculate orbital correlations defined by 
$C\equiv \sum_{ij}^{\prime}\langle W_i W_j \rangle$, 
where $W_i$ denotes a psuedo-spin
operator of $3 x^2 - r^2$ or $3y^2 - r^2$ 
($W_i \equiv -\frac{1}{2} T_i^z +(-) \frac{\sqrt{3}}{2} T_i^x$ 
for $3x^2(y^2)-r^2$, and 
$T_i^{\mu}=\frac{1}{2}\sum_{\nu\nu^{\prime}} 
\hat{\sigma}_{\nu\nu^{\prime}}^{\mu} 
c^{\dagger}_{i\nu}c_{i\nu^{\prime}}$ 
with the Pauli matrix $\hat{\sigma}_{\nu\nu^{\prime}}^{\mu}$).  
Prime at the sum means that $i$ and $j$ run over one sublattice. 
The possible patterns on the lattice are illustrated 
in Fig.\ref{fig3}. 
The results reveal that $C$ for (a) is the largest\cite{values_of_C}. 
The pattern (a) agrees with the experimental indication\cite{Jirak}. 
Thus, 
the basic orbital structure in charge-ordering phenomena is 
determined irrespective of magnetic structure 
due to difference of energy scales. 
This situation is captured well within the model 
(\ref{Mn-Hamiltonian}) 
in which the spin degrees of freedom are frozen out.  

Finally, it is worth mentioning effects 
of the three dimensionality neglected above.  
In 2D, there happens the imbalance of populations 
in $d_{x^2 - y^2}$ and $d_{3 z^2 - r^2}$ as described above. 
In 3D case, a hopping along $z$-axis would 
reduce the imbalance. 
We note that, in the NI case, 
such a reduction of OP in 3D from that in 2D 
decreases the coherence as inferred from the comparison 
between 3D case in ref. \ref{Shiba_Shiina_Takahashi} 
and 2D one in Fig.\mbox{\,}\ref{fig1}.  
This is simply because the charge incoherence can not be induced 
if the orbital is completely polarized. 
The orbital depolarization may also be simulated 
by introducing a chemical potential difference 
between the two orbitals to compensate the existing OP. 
We have performed calculations 
for the $\sqrt{10}$\mbox{\,}$\times\!\sqrt{10}$ -site 2D Hamiltonian 
with such a chemical potential difference. 
The results show that the Drude weight becomes more suppressed. 
For example, if the chemical potential difference is tuned to keep 
the vanishing OP, 
$\frac{D}{N_{\rm eff}}$\mbox{\,}$\sim$\mbox{\,}0.45 at $\delta$=0.2 
as compared to 
$\frac{D}{N_{\rm eff}}$\mbox{\,}$\sim$\mbox{\,}0.67 
in the case of Fig.\mbox{\,}\ref{fig1}. 
The broad shape of incoherence within the LH band 
near the Mott transition is found to be qualitatively unchanged 
after the chemical potential is introduced. 
These suggest the possibility that 
the depolarization due to the three dimensionality would also 
make the charge transport more incoherent. 

  In summary, we have investigated the optical conductivity 
$\sigma (\omega)$ of the double-exchange model 
with both the $e_{g}$-orbital degeneracy 
and the Coulomb interaction. 
Assuming the spin polarization and 
the strong Hund's-rule coupling in the 2D system, 
we have examined the effects of the orbital degeneracy and 
the on-site interaction.  
We have obtained in $\sigma (\omega)$ 
a suppressed Drude weight and a mid-gap incoherence 
with a broad structure 
in the metallic state near the Mott transition. 
The results reproduce to a considerable extent 
the incoherent charge dynamics observed experimentally in Mn oxides. 
On a quantitative level, however, 
the calculated Drude weight does not appear 
to be sufficiently small\cite{Okimoto_let,Okimoto_reg}. 
Effects of fluctuations of JT distortion 
would be an additional source of the incoherence, which 
is an interesting subject of further studies 
for more quantitative comparison. 
We have also found an anomalous feature in charge transport 
and the charge correlation 
at quarter filling, which reproduces basic structure of 
the orbital and charge ordering observed in experiments. 

The authors thank T. Ogitsu and S. Todo for useful advice
to develop parallel-processing programs.
Y.M. is supported by Research Fellowships of Japan Society
for the Promotion of Science (JSPS) for Young Scientists.
This work is supported by `Research for the Future Program' 
from JSPS (JSPS-RFTF 97P01103).
A part of the computations was performed using the
facilities of the Supercomputer Center,
Institute for Solid State Physics (ISSP), University of Tokyo.
Parallel calculations were done
in the supercomputer FUJITSU VPP500 in ISSP
as well as in the VPP700 in Kyushu University.

\begin{figure}[h]
\caption{Doping dependences of (a) the total weight $K$, 
(b) the effective carrier density $N_{\rm eff}$ and 
the Drude weight $D$ at $U/t_0$=16. 
For comparison, gray lines show results in the non-interacting case,  
where the total weight, the effective carrier density 
and the Drude weight in the thermodynamic limit 
$N_{\rm s}$\mbox{\,}$\rightarrow$\mbox{\,}$\infty$ 
are shown by the dotted line in (a),  
dotted one in (b) and solid one in (b), respectively.
}
\label{fig1}
\end{figure}

\begin{figure}[h]
\caption{Incoherent part of the optical conductivity at $U/t_0=16 $
for (a) the half-filled case in the 10-site system 
under the anti-periodic BC, 
(b) 2 holes in the 4$\times$4 sites 
under the mixed BC,  
(c) 4 holes in the 4$\times$4 sites 
under the anti-periodic BC.  
Inset in (b) shows the non-interacting case at the same filling and 
the same BC. 
Delta functions are broadened with width of $
0.05 t_0$. 
We choose the BC to realize 
the lowest-energy ground state. 
}
\label{fig2}
\end{figure}

\begin{figure}[h]
\caption{Doping dependences of the charge correlation. 
An anomaly at $\delta$=0.5 for the model (\ref{Mn-Hamiltonian}) 
is obtained while only the monotonic behavior is seen 
in the usual Hubbard model. 
Insets (a), (b) and (c) display possible staggered patterns 
of orbitals at $\delta$=0.5 in 4$\times$4 cluster. 
Simple circles and anisotropic symbols denote Mn$^{4+}$ and Mn$^{3+}$ 
with $3x^2-r^2$/$3y^2-r^2$, respectively. 
}
\label{fig3}
\end{figure}


\begin{thebibliography}{99}
\bibitem{Ramirez_rev} A. P. Ramirez: 
J. Phys. Condens. Matter {\bf 9} (1997) 8171 and references therin. 
\bibitem{Zener} C. Zener: Phys. Rev. {\bf 82} (1951) 403.
\bibitem{Anderson_Hasegawa} P.\mbox{\,}W.\mbox{\,}Anderson and 
H.\mbox{\,}Hasegawa: Phys.\mbox{\,}Rev. {\bf 100} (1955) 675. 
\bibitem{de_Gennes} P. G. de Gennes: Phys. Rev. {\bf 118} (1960) 141. 
\bibitem{Kubo_Ohata} K. Kubo and N. Ohata: 
J. Phys. Soc. Jpn. {\bf 33} (1972) 21. 
\bibitem{Frkw_DMF} N. Furukawa: J. Phys. Soc. Jpn. {\bf 63} (1995) 3214. 
\bibitem{Okimoto_let} Y. Okimoto, T. Katsufuji, T. Ishikawa, A. Urushibara, 
T. Arima and Y. Tokura: Phys. Rev. Lett. {\bf 75} (1995) 109.
\bibitem{Okimoto_reg} Y. Okimoto, T. Katsufuji, T. Ishikawa, 
T. Arima and Y. Tokura: Phys. Rev. B {\bf 55} (1997) 4206.
\bibitem{Millis}\label{Millis} 
A.\mbox{\,}J.\mbox{\,}Millis, R.\mbox{\,}Mueller 
and B.\mbox{\,}I.\mbox{\,}Shraiman: 
Phys.\mbox{\,}Rev.\mbox{\,}B {\bf 54} (1996) 5405. 
\bibitem{Shiba_Shiina_Takahashi} \label{Shiba_Shiina_Takahashi}
H.\mbox{\,}Shiba, R.\mbox{\,}Shiina and A.\mbox{\,}Takahashi: 
J.\mbox{\,}Phys.\mbox{\,}Soc.\mbox{\,}Jpn. {\bf 66} (1997) 941. 
\bibitem{Takahashi_Shiba} 
A. Takahashi and H. Shiba: 
Euro. Phys. J. B {\bf 5} (1998) 413.
\bibitem{Brito_Shiba} 
P. E. de Brito and H. Shiba: 
Phys. Rev. B {\bf 57} (1998) 1539.
\bibitem{Ishihara_Yamanaka_Nagaosa} S. Ishihara, M. Yamanaka and N. Nagaosa
: Phys. Rev. B {\bf 56} (1997) 686. 
\bibitem{Horsch_finite_temp}\label{Horsch_finite_temp}
 P.\mbox{\,}Horsch, J.\mbox{\,}Jaklic 
and F.\mbox{\,}Mack: 
Phys.\mbox{\,}Rev.\mbox{\,}B {\bf 59}\mbox{\,}(1999)\mbox{\,}6217.
\bibitem{Woodfield} B.F.Woodfield, M.L.Wilson and J.M.Byes: 
Phys.\mbox{\,}Rev.\mbox{\,}Lett. {\bf 78} (1997) 3201. 
\bibitem{Horsch_temperature}
In ref.\ref{Horsch_finite_temp}, 
the authors presented their data only above the temperature of 
$T=0.3 t_0$, where $t_0$ is a hopping amplitude which 
the authors themselves estimated to be 0.2 eV. 
\bibitem{Motome_Imada} \label{Motome_Imada}
Y. Motome and M. Imada: 
J. Phys. Soc. Jpn. {\bf 68} (1999) 16. 
\bibitem{Ishikawa} T. Ishikawa, T. Kimura, T. Katsufuji and Y. Tokura: 
Phys. Rev. B {\bf 57} (1998) R8079.    
\bibitem{NEC_Motome_Nakano_Imada} \label{NEC_Motome_Nakano_Imada}
Y. Motome, H. Nakano and M. Imada: 
preprint cond-mat/9811221. 
\bibitem{Motome_reg} \label{Motome_reg}
Y. Motome and M. Imada: preprint cond-mat/9903183. 
\bibitem{frac_expansion} E.\mbox{\,}Gagliano and 
C.\mbox{\,}Balseiro: 
Phys.\mbox{\,}Rev.\mbox{\,}Lett. {\bf 59} (1987) 2999.
\bibitem{Nakano_Imada_usual-Hub} H.\mbox{\,}Nakano and M.\mbox{\,}Imada: 
J.\mbox{\,}Phys.\mbox{\,}Soc.\mbox{\,}Jpn. 
{\bf 68} (1999) 1458. 
\bibitem{comment_quarter_filling_boundary}
At this filling, the optimized boundary 
of $4\times4$-site cluster at $U/t_0=16$ is a mixed one. 
\bibitem{comment_free_case} For comparison, the corresponding quantity 
for the non-interacting case is 0.25. 
\bibitem{Kuwahara} H. Kuwahara, Y. Tomioka, A. Asamitsu, Y. Moritomo
and Y. Tokura: Science {\bf 270} (1995) 961.
\bibitem{Jirak} Z. Jirak, S. Krupicka, Z. Simsa, M. Dlouha 
and S. Vratislav: J. Magn. Magn. Mater. {\bf 53} (1985) 153.
\bibitem{Tomioka} Y. Tomioka, A. Asamitsu, Y. Moritomo, H. Kuwahara 
and Y. Tokura: Phys. Rev. Lett. {\bf 74} (1995) 5108.
\bibitem{Ishihara_CO}
\label{Ishihara_CO} 
S.\mbox{\,}Ishihara and N.\mbox{\,}Nagaosa : 
J.\mbox{\,}Phys.\mbox{\,}Soc.\mbox{\,}Jpn. {\bf 66} (1997) 3678. 
\bibitem{Anisimov} V. I. Anisimov, I. S. Elfimov and 
K. Terakura: Phys. Rev. B {\bf 55} (1997) 15494.
\bibitem{Mizokawa} T. Mizokawa and A. Fujimori: 
Phys. Rev. B {\bf 56} (1997) R493.
\bibitem{murakami} Y. Murakami, H. Kawada, H. Kawata, M. Tanaka, 
T. Arima, Y. Moritomo and Y. Tokura: 
Phys. Rev. Lett. {\bf 80} (1998) 1932. 
\bibitem{values_of_C} $C\sim1.67$ for (a), $\sim$1.41 for (b), 
$\sim$1.29 for (c) of $3x^2-r^2$ and 
$\sim$1.49 for (c) of $3y^2-r^2$ 
when the mixed BC is of $\Phi=(\pi,0)$. 
\end{thebibliography}
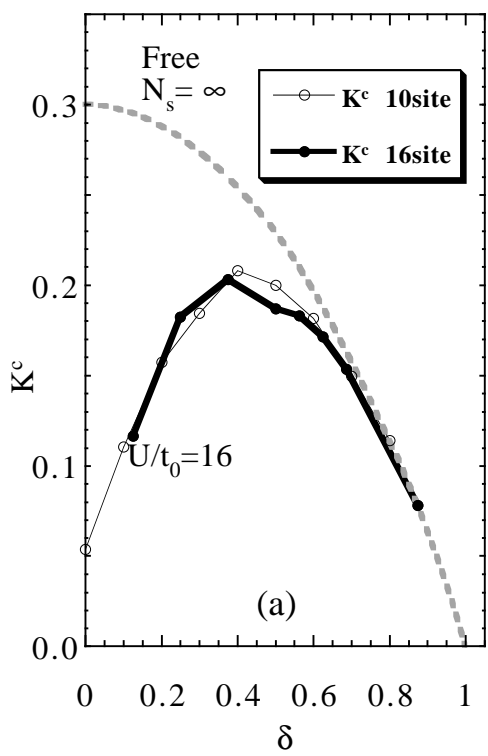
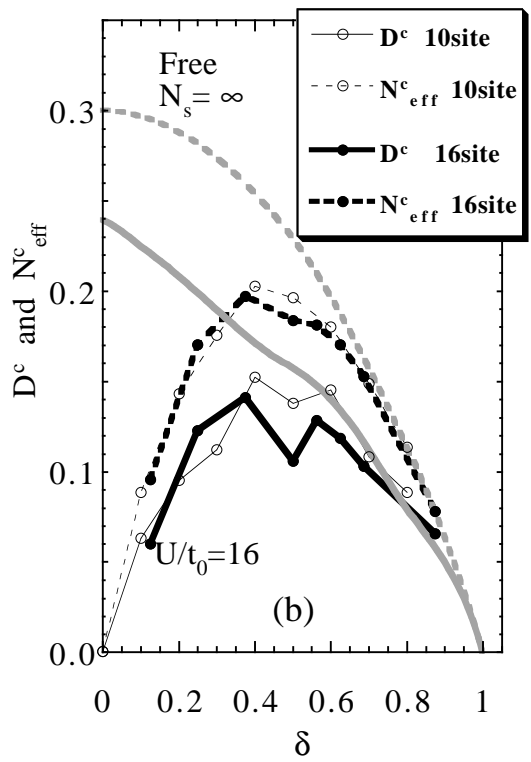
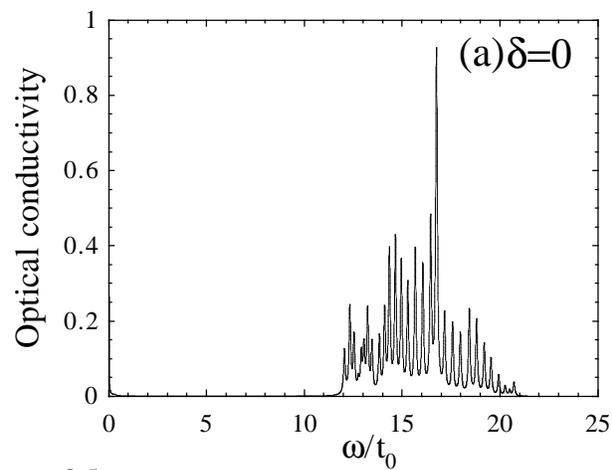
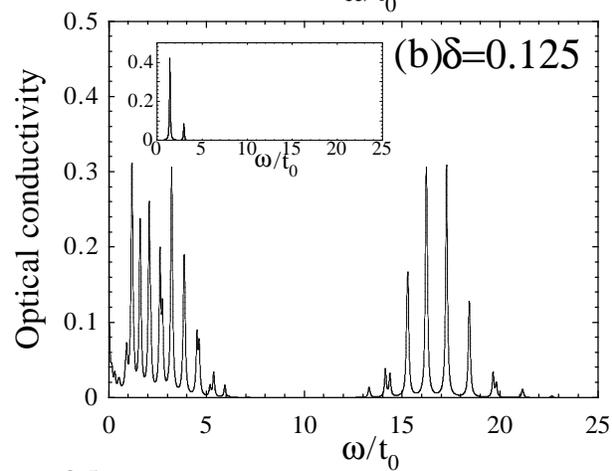
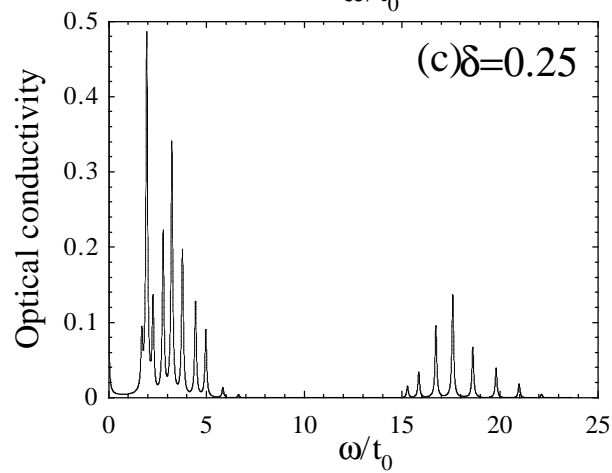
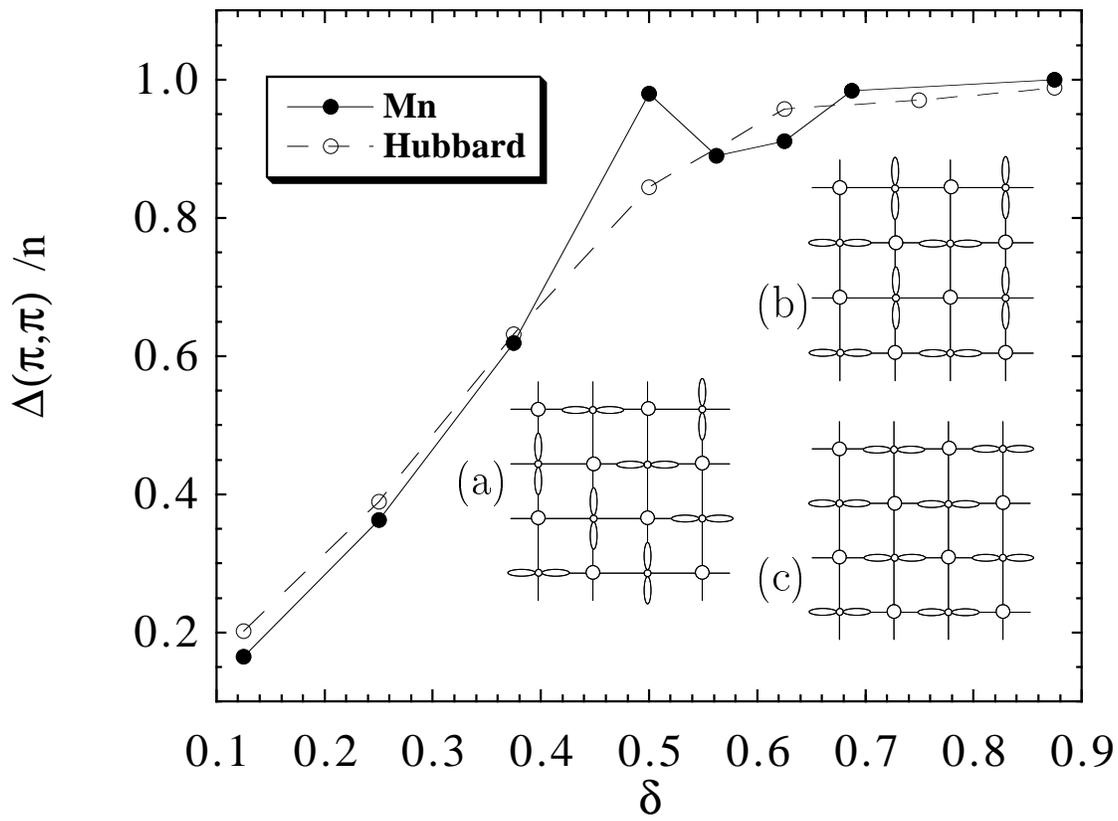
\end{document}